# Stimulated emission tomography for entangled photon pairs with different detection spectral range

YIQUAN YANG, PEIYU ZHANG, AND XIAO-SONG MA*

*National Laboratory of Solid State Microstructures, School of Physics and Collaborative Innovation Center of Advanced Microstructures, Nanjing University, Nanjing, China*
*xiaosong.ma@nju.edu.cn*

**Abstract:** Frequency non-degenerate entangled photon pairs have been employed in quantum communication, imaging, and sensing. To characterize quantum entangled state with long-wavelength (infrared, IR or even terahertz, THz) photon, one needs to either develop the single-photon detectors at the corresponding wavelengths or use novel tomography technique, which does not rely on single-photon detections, such as stimulated emission tomography (SET). We use standard quantum state tomography and SET to measure the density matrix of entangled photon pairs, with one photon at 1550 nm and the other one at 810 nm, and obtain highly consistent results, showing the reliability of SET. Our work paves the way for efficient measurement of entangled photons with highly dissimilar frequencies, even to the frequencies where single-photon detections are not available.

## 1. Introduction

Entanglement is the quintessential resource in emerging quantum technology such as quantum computation [1], quantum communication [2] and quantum metrology [3]. Characterizing an entangled state is a prerequisite for many quantum information tasks. For the photonic system, quantum state tomography (QST) [4] can provide the density matrix by using coincidence measurement with single-photon counting modules (SPCM). Silicon-based SPCM (Si-SPCM) has been widely used for detecting visible photons. Single-photon detector technology makes great improvements in recent years and extends the detection spectral range into near-infrared (NIR) and IR regions [5]. For instance, one can use the frequency up-conversion detector (FUCD) [6,7] and superconducting single-photon detectors (SSPDs). Although both offer excellent performance, extra assistant equipment is needed. For FUCD, an auxiliary pump laser and an extra nonlinear crystal are required to facilitate high-efficiency single-photon frequency conversion. For SSPD, it requires cryogenic operation environment. To develop a simpler detection scheme, particularly for entangled state characterization, is highly desirable.

Stimulated emission tomography (SET), proposed by Liscidini et al. [8], is an alternative and simple method to characterize the quantum entangled state. It doesn't require single-photon counting and coincidence measurements as compared with standard QST. Stimulated emission in non-linear optics is realized by injecting particularly prepared seed laser into a non-linear crystal. In [9], Fang et al. first experimentally demonstrated the SET method by measuring the joint spectral density of entangled photon pairs, in which they used laser light and the classical light detector. By employing carefully prepared seed laser with different degrees of freedom, such as wavelength [9,10], polarization [11-13], and path [13], one can characterize different correlated/entangled states of photon pairs.

All of the previous experimental demonstrations of SET concentrate on entangled photon pairs in a single detection spectral range [11-13]. Both signal and idler photons can be detected by the single-photon detectors made by the same materials, Si. However, highly non-degenerate photon pairs could have unique advantages in various quantum protocols. For instance, one could use visible photons for interfacing quantum memories and telecom photons for long-distance communications [14]. In this scenario, one needs different detection technologies with different detection spectral ranges, e.g. Si-SPCM and SSPD. By using SET in this scenario, one can avoid using SSPD and hence greatly reduce the complexity of the characterization step of the experiment. Moreover, to extend SET to polarization entanglement with 1550 nm photons may enable new protocols involving long-distance communications. In this work, we investigate frequency non-degenerate polarization-entangled state, consisting of one photon at 810 nm and another one at 1550 nm. Such entangled photon pairs have been used in practical entanglement-based quantum key distribution experiments, which allows the 810 nm photon to be detected locally with a low-noise Si-SPCM at Alice's node and the 1550 nm photon to be transmitted to Bob's site via a low-loss optical fiber [15]. Frequency non-degenerate entangled photons have the application prospect in quantum imaging [16] and sensing [17-19]. Classical IR imaging/sensing technology has wide applications in material characterization, gas sensing, biological research. IR imaging technology with the help of quantum resource could bring revolutions in the field. Therefore, SET provides an alternative quantum state tomography solution, which is helpful for some spectral ranges without efficient single-photon detectors. In short, our work has following several novelties compared with previous works: 1. Our work is the first demonstration of SET across different wavelength ranges categorized by the detection capabilities of single-photon detectors; 2. We extend SET to polarization entanglement with 1550 nm photons, which may enable new protocols involving long-distance communications and stimulated emission.

## 2. Experiment

In our experiment, we prepare entangled state via type-I spontaneous parametric down-conversion (SPDC) process in non-linear crystal (NLC) 5% MgO·LiNbO3. A single-mode laser centered at 532 nm incident on NLC can generate a pair of correlated photons with central wavelength $\lambda_s$ = 810 nm and $\lambda_i$ = 1550 nm, i.e. $|532\rangle_p \to |810\rangle_s |1550\rangle_i$, where indexes $s$ and $i$ denote signal and idler photons. The experimental setup, depicted in Fig. 1, is composed of three parts which are entangled state preparation, tomography and seed state preparation. The pump light with an average power of 50 mW is prepared in diagonal polarization. It is then separated by a beam displacer (BD-1) into horizontal polarized light in upper path and vertical polarized light in lower path which will be rotated to horizontal polarization by D-shaped half-wave plate (D-HWP). Horizontal polarized pump light in both arms is focused onto NLC, which has a thickness of 2 mm and its optics axes was cut at $68°$ to the surface for type-I SPDC phase matching condition. This process generates photon pairs both in vertical polarization direction. The signal and idler photon are separated by Dichroic Mirror (DM) into transmission and reflection directions, respectively. Correlated photon pair generated in lower path is rotated by a D-HWP to horizontal direction, i.e. the polarization state of correlated photon pairs is $|H\rangle_s |H\rangle_i$ and $|V\rangle_s |V\rangle_i$ in lower path and upper path respectively. The signal and idler photon generated in both arms are combined at the end of BD-2 and BD-3 separately, then the resultant state is $1/\sqrt{2}(|H\rangle_s |H\rangle_i + e^{i\theta}|V\rangle_s |V\rangle_i)$, where $\theta$ is the relative phase of photons propagating along upper and lower path. The phase $\theta$ can be adjusted via appropriate tilting of BD-2 and BD-3 in our experiment.

The nonlinear optical process facilitates SET is different frequency generation (DFG), which obey to phase matching and energy conservation conditions, similar as those in SPDC. By injecting seed light (~ 810 nm), the stimulated generated light (~1550 nm) can be easily detected with power meter. As mentioned above, we use polarization entangled photon pairs. By properly setting the polarization of seed light, we can control the efficiency of DFG. Based on the measurement result of stimulated idler photon detection under different polarized seed light, we can analyze entangled state. In our experiment we use power meter for idler photons detection (1550 nm). Additionally, we also use Optical Spectrum Analyzer (OSA) to measure idler photons' spectrum. Compared with QST employing high performance single-photon detectors, SET only requires visible light laser and classical infrared light detector, which are easier and faster to obtain. In SET approach, we set the seed light appropriately at 810 nm and incident onto the NLC to generate stimulated idler photons via DFG, which can be detected by classical intensity detector.

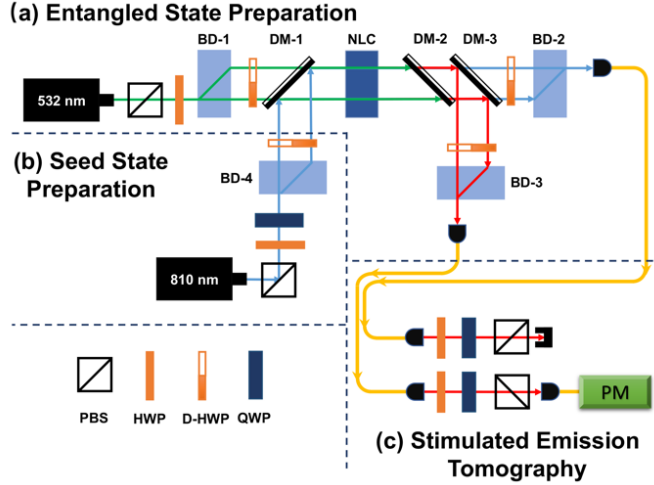

Fig. 1. Schematic of the experimental setup for stimulated emission tomography (SET). Three modules include (a) entangled state preparation, (b) seed state preparation and (c) stimulated emission tomography. The entangled state preparation module generates the polarization-entangled state of 810 nm and 1550 nm photons. The details can be found in the main text. Note that DM-1 is used to combine pump and signal light; DM-2 filters out the pump light; and DM-3 is used to separate signal and idler photons. The seed state preparation module, prepares the seed light centered at 810 nm with different polarization states, which mimics the signal photon that would be generated spontaneously. The polarization of the seed beam exiting from this module is controlled by a combination of HWP, QWP, and BD. The tomography module is used to measure down-converted photons for quantum state tomography. In SET, we only need to detect stimulated generated 1550 nm light with a power meter (PM). BD: Beam Displacer; NLC: Non-linear Crystal; DM: Dichroic Mirror; HWP: Half Wave Plate; D-HWP: D-shaped Half Wave Plate; QWP: Quarter Wave Plate; PBS: Polarized Beam Splitter.

To obtain the density matrix with SET, different polarized states of the seed light are prepared in the seed state preparation module (Fig. 1**b**) from a continuous wave laser (SolsTis 3500 XL SRX, M SQUARED). Then they are sent into NLC, which allows idler photon generation via DFG. We prepare seed laser in six different polarization states (i.e. $|H\rangle, |V\rangle, |D\rangle, |A\rangle, |R\rangle, |L\rangle$) and analyze the stimulated idler photon by single-photon quantum state tomography. This is equivalent to make coincidence measurements in 36 settings which can provide us full knowledge about the density matrix of the entangled state [20].

Note that because we employ Type-I SPDC phase matching condition for both arms in NLC, i.e. $|H\rangle_p \rightarrow |V\rangle_s |V\rangle_i$, we need rotate H polarization component of superposition seed states (i.e. $|D\rangle, |A\rangle, |R\rangle, |L\rangle$) into V to stimulate idler photon generation. Then the diagonal polarization state of the seed light, $1/\sqrt{2}(|H\rangle_s^{upper} + |V\rangle_s^{lower})$, should be transformed into $1/\sqrt{2}(|V\rangle_s^{upper} + |V\rangle_s^{lower})$ by using D-HWP, which is a superposition state of vertical polarized state in upper arm and lower arm. The incident seed light intensity is about 20 mW with the central wavelength of $\lambda_s = 810$ nm. We obtained a stimulated idler output power is approximately 0.06 nW at the maximum.

## 3. Results

To benchmark the SET result, we first perform standard QST with coincidence measurements with single-photon detectors. By adjusting path delay of signal photons, the quantum state is close to the maximally entangled state:

$$|\Psi\rangle = \frac{1}{\sqrt{2}}(|H\rangle_s|H\rangle_i + |V\rangle_s|V\rangle_i). \tag{1}$$

In Fig. 2**a**, we show the real and imaginary parts of the reconstructed density matrix of $|\Psi\rangle$, from which we obtain the state fidelity of 96.660±0.001% with respect to the state of Eq. (1), and we also calculate the concurrence [21] which is 93.409±0.001%.

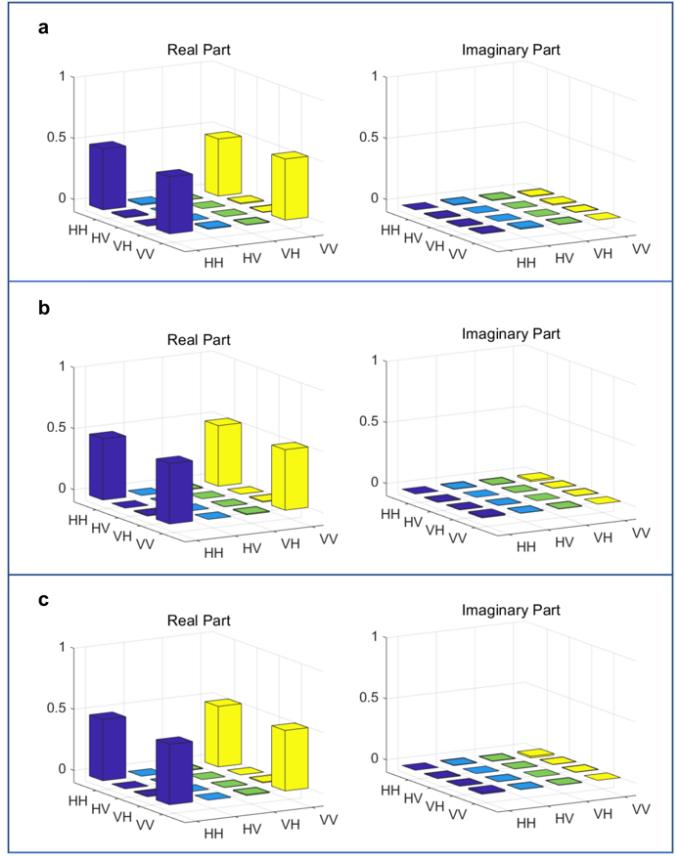

Fig. 2. Quantum state tomography (QST) and stimulated emission tomography (SET) measurement results are shown in panel (a) and (b)-(c), respectively. Left and right figure is real and imaginary part of polarization-entangled state density matrix. (a): the signal and idler photon is detected by a Si Single-Photon Count Module and a Superconducting Single-Photon Detector (SSPD). (b) and (c): the stimulated idler photon is detected by power meter and Optical Spectrum Analyzer (OSA) individually.

We then proceed to the results of SET, which requires the 'Seed state preparation' module, as shown in Fig. 1**b**. A combination of HWP, QWP and BD prepares the required quantum states of the seed exiting from the module. Our experimental procedure is the following: By rotating fast-axis angles of HWP and QWP in module (b), we can control the polarization of seed photon up to a minor birefringent phase caused by the optical elements in the interferometer. We tilt BD-4 to compensate this unwanted phase. By doing so, we prepare the required seed photons for SET. This process is a polarization measurement.

During the experiment, we also use a power meter to measure the seed power for each polarization measurement of the seed. Via analyzing stimulated idler power data with the maximum likelihood technique, same as QST, we can construct the density matrix shown in Fig. 2**b**, and obtain the fidelity of 99.381±0.012% with respect to the state of Eq. (1) and the concurrence is 98.812±0.022%. Moreover, we also use an OSA to measure the power of idler along with its spectrum. The results are shown in Fig. 2**c** and we obtain the state fidelity of 99.575±0.001% with respect to the state of Eq. (1) and the concurrence is 99.673±0.001%. As shown in Fig. 2**b** and 2**c**, we have obtained quite consistent tomography results from power meter and OSA, which imply the robustness and reliability of the SET technique.

Although the density matrix of the polarization-entangled state acquired by QST and SET are quite close, there exists a small discrepancy in the relative phase between $|H_s\rangle|H_i\rangle$ and $|V_s\rangle|V_i\rangle$. The value of $\theta$ inferred by QST is 0.0138, whereas the value evaluated by SET based on OSA and power meter measurement results is 0.0247 and 0.0252 respectively. The reasons for this small discrepancy (~0.01) are that we have different spectral distribution for the idler photons collected in QST and SET. In QST, the measured phase between $|H_s\rangle|H_i\rangle$ and $|V_s\rangle|V_i\rangle$ is average of all spectra that satisfy the phase-matching condition of SPDC. Note that in ref [11], this phase discrepancy is as large as 0.289, which is about 30 times larger than our results. This could mainly because we use a colinear configuration whereas in ref [11] the authors used non-colinear configuration.

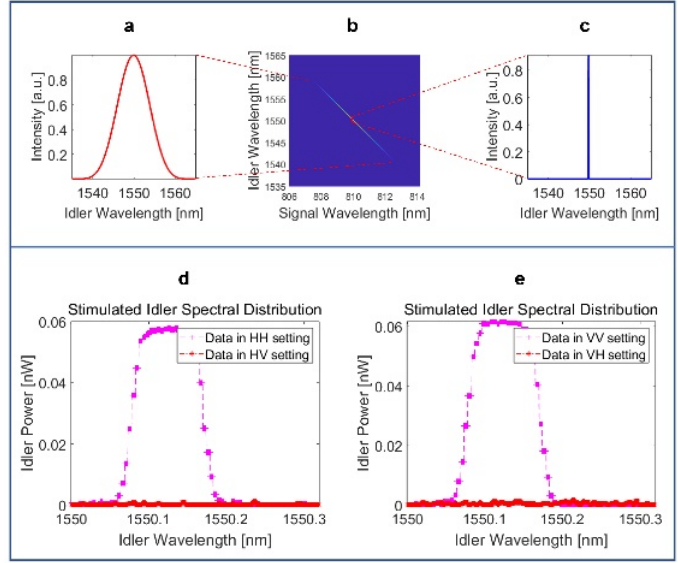

Fig. 3. The joint spectral intensity profile of correlated photon pairs is shown in (b). The left figure (a) display spectral distribution of idler photon from SPDC. The right figure (c) shows idler spectral distribution generated by a stimulated process with narrow continuous seed light at the wavelength of 810 nm in DFG. The figure (d) and (e) are stimulated idler spectral distribution measured by Optical Spectrum Analyzer in four measurement settings. Limited by paper space, stimulated idler spectral distributions demonstrate above only include four measurement settings.

Another discrepancy of the results obtained from QST and SET is the state fidelity to Eq. (1). We obtain higher fidelities by using SET as compared to QST. This discrepancy could be explained by the frequency-polarization correlations [12,22]. In a typical SPDC source for generating polarization-entangled photon pairs, photons' polarization and frequency are correlated due to the phase-matching condition and pump spectrum. To obtain high-quality polarization-entangled states, one normally uses narrow-band interference filters to select partial bandwidth of the photons to approximate a single frequency mode [23]. By doing so, the frequency uncorrelated photon pairs will be closer to the maximally entangled states in polarization. Here we use a narrow-band seed laser to stimulate the generation of idler photons. Therefore, idler photons generated via DFG, which fulfill the phase-matching condition given by the narrow-band seed light, have significantly narrower bandwidth as compared to that generated via SPDC. This could be viewed as an active filtering process and hence the state fidelity to the maximally entangled state is higher in the case of SET. In Fig. 3**b**, we plot the calculated joint spectral intensity (JSI) profile of our SPDC process. The spectrums of idler photons generated from SPDC and DFG are shown in Fig. 3**a** and Fig. 3**c**, respectively. It is clear to see that the spectrum of SPDC is much wider than that of DFG. In Fig. 3**d** and Fig. 3**e**, we show stimulated idler spectrum detected by OSA in four measurement settings. Limited by spectral resolution of OSA, the measured FWHM of the stimulated idler spectrum is larger than the theoretical prediction shown in Fig. 3**c**. With narrow-band seed SET, we could set an upper bound of the state fidelity for the entangled photon pairs generated from SPDC. Note it is difficult and impractical to use QST with the narrow-band filtering on the single photons because the count rate of single photons with narrow-band filtering (~100 kHz in our case) becomes prohibitively low. Using narrow-band seed SET, we could probe the limit of the fidelity set by the bandwidth filtration. We believe this is a valid and useful technique for developing high- quality entangled photon-pair sources, which require optimizations over many degrees of freedom, including the bandwidth of the wavelength, polarization, spatial modes, etc. Note that the state fidelity of SET results with respect to that of QST is 97.6707% and 97.9301% based on data collected by OSA and PM, which implies the reliability of the SET approach. To further study the origin of the discrepancy in fidelity, one may use energy-resolved polarization SET by scanning the seed's wavelength across the range fulfilling phase matching conditions, as shown in ref [12].

## 4. Conclusion

In this work, we demonstrate the SET method is reliable for entangled photon pairs with 740 nm wavelength separation, i.e. one photon in the visible range ($\lambda_s = 810$ nm) and the other one in the near-infrared range ($\lambda_i = 1550$ nm). The infrared spectral region that the idler located in possesses pervasive application in quantum technology, such as quantum communication, and remote sensing. We show that SET can be an alternative tomography method without requiring a single-photon detector. Especially, employing SET is quite beneficial in non-degenerate entangled photon pairs tomography. We expect this method could be helpful to the unexplored spectral range in quantum technology, such as mid-infrared and far-infrared regions [18,24], or even tera-hertz spectral range [25,26].

## 5. Acknowledgments

This research is supported by the National Key Research and Development Program of China (2017YFA0303700, 2019YFA0308704), National Natural Science Foundation of China (Grant No. 11674170, 11621091, 11321063, 11690032), NSF Jiangsu Province (No. BK20170010), BRICS STI Framework Programme (No. 61961146001), the program for Innovative Talents and Entrepreneur in Jiangsu, and the Fundamental Research Funds for the Central Universities.